\documentclass[preprint]{imsart}%
\usepackage{amsfonts}
\usepackage{amsmath}
\usepackage{amssymb}
\usepackage{graphicx}%
\providecommand{\U}[1]{\protect\rule{.1in}{.1in}}
\RequirePackage[OT1]{fontenc}
\RequirePackage{amsthm,amsmath}
\RequirePackage[numbers]{natbib}
\RequirePackage[colorlinks,citecolor=blue,urlcolor=blue]{hyperref}
\arxiv{arXiv:1708.05341}
\startlocaldefs
\numberwithin{equation}{section}
\theoremstyle{plain}

\endlocaldefs
\begin{document}
\begin{frontmatter}
\title{An Approximate Likelihood Perspective on ABC Methods}
\runtitle{Approximate Likelihood for ABC}
\begin{aug}
\author{\fnms{George} \snm{Karabatsos}\thanksref{t1}\ead[label=e1]{gkarabatsos1@gmail.com}}
\and
\author{\fnms{Fabrizio} \snm{Leisen}\thanksref{t2}\ead[label=e2]{fabrizio.leisen@gmail.com}}
\ead[label=e3]{}
\ead[label=e4]{}
\thankstext{t1}{Corresponding author. The research of George Karabatsos was supported by NSF grant SES-1156372.}
\thankstext{t2}{The research of Fabrizio Leisen was supported by the European Community's Seventh Framework Programme [FP7/2007-2013] under grant agreement no. 630677.}
\runauthor{G. Karabatsos and F. Leisen}
\affiliation{University of Illinois-Chicago and University of Kent}
\address{George Karabatsos\\
Departments of\\
Educational Psychology\\
and Mathematics, Statistics, \\
and Computer Science\\
1040 W. Harrison St. (MC 147)\\
Chicago, IL 60607\\
\printead{e1}\
\phantom{\ }\printead*{e3}}
\address{Fabrizio Leisen\\
School of Mathematics, Statistics,\\
and Actuarial Sciences\\
Cornwallis Building\\
University of Kent\\
Canterbury, CT2 7NF\\
United Kingdom\\
\printead{e2}\
\phantom{\ }\printead*{e4}}
\end{aug}
\begin{abstract}
We are living in the big data era, as current technologies and networks allow for the easy
and routine collection of data sets in different disciplines. Bayesian Statistics offers a flexible
modeling approach which is attractive for describing the complexity of these datasets.
These models often exhibit a likelihood function which is intractable due to the large
sample size, high number of parameters, or functional complexity. Approximate
Bayesian Computational (ABC) methods provides likelihood-free methods for
performing statistical inferences with Bayesian models defined by intractable likelihood
functions. The vastity of the literature on ABC methods created a need to review and
relate all ABC approaches so that scientists can more readily understand and apply
them for their own work. This article provides a unifying review, general representation,
and classification of all ABC methods from the view of approximate likelihood theory.
This clarifies how ABC methods can be characterized, related, combined, improved,
and applied for future research. Possible future research in ABC is then suggested.
\end{abstract}
\end{frontmatter}

\section{Introduction}

Bayesian models are applied for statistical inference in many scientific
fields. The posterior distribution is the main object of Bayesian Statistics
and it is the result of the combination of two information sources, namely the
prior distribution, which reflects extra-experimental knowledge, and the
likelihood function, which formalizes the information provided by the data
through the use of a given statistical model.

Posterior inferences with such models can be undertaken by applying classical
Monte Carlo (MC)\ methods, which provide iterative algorithms that can
generate approximate samples from the posterior distribution, without the
marginal likelihood. They include Markov chain Monte Carlo (MCMC) methods,
such as Metropolis-Hastings (MH), Gibbs, slice, and adaptive rejection
sampling algorithms
\citep[e.g.,][]{RobertCasella04,BrooksGelmanJonesMeng11}%
; and include population MC
\citep[PMC, e.g.,][]{CappeEtAl04}%
, sequential MC
\citep[SMC, e.g.,][]{DelMoralEtAl06}%
), and other importance sampling (IS) methods
\citep[e.g.,][]{Liu01}%
. Alternatively, variational inference (VI)\ optimization methods can be
applied to find a tractable density function that minimizes the divergence to
the exact posterior density
\citep[][]{BleiEtAl17}%
.

However, modern \textquotedblleft big data\textquotedblright\ applications
require complex models and demanding computational techniques; in some of
these situations, MCMC or VI methods may be extremely slow or even impossible
to implement. ABC methods are useful in the general scenario where for the
given Bayesian model of interest for data analysis, the likelihood is not
easily evaluated or intractable, but it is still possible to either draw
samples from this likelihood conditionally on the model parameters
\citep[e.g.,][]{FuLi97,WeissVonHaeseler98,PritchardEtAl99,BeaumontEtAl02}%
; or to find a point-estimate of some model parameter function based on a
sufficient statistic of the data
\citep[e.g.,][]{TavareEtAl97}%
, on an empirical likelihood
\citep{MengersenPudloRobert13}%
, or on a bootstrap method
\citep[e.g.,][]{ZhuMarinLeisen16}%
.

Approximate Bayesian Computational (ABC) methods provide \textquotedblleft
likelihood-free\textquotedblright\ methods for performing statistical
inferences with Bayesian models defined by intractable likelihood functions.

Since ABC was introduced
\citep[][]{TavareEtAl97,PritchardEtAl99,BeaumontEtAl02}%
, it has been applied to many scientific fields, models, and problems
involving intractable likelihoods. ABC has been applied in independent and
related fields of archaeology
\citep[e.g.,][]{WilkinsonTavare09,TsutayaYoneda13,CremaEtAl14}%
; astronomy and cosmology
\citep
[e.g.,][]{LewisBridle02,CameronPettitt12,WeyantEtAl13,AkeretEtAl15,KacprzakEtAl18}%
; various biology subfields including cell biology
\citep[e.g.,][]{JohnstonEtAl14,VoEtAl15mela,VoEtAl15quant}%
, ecology and molecular ecology
\citep
[e.g.,][]{ToniEtAl09,WilkinsonTavare09,Beaumont10,IlvesEtAl10,Wood10b,LagarriguesEtAl15,BennettEtAl16,FasioloWood18}%
, evolutionary biology and molecular ecology
\citep
[e.g.,][]{HickersonEtAl06,JakobssonEtAl06,FagundesEtAl07,RosenblumEtAl07,VerduEtAl09,Beaumont10,FanKubatko11,LombaertEtAl11,RatmannEtAl12,SlaterEtAl12,EstoupEtAl18}%
, genetics
\citep
[e.g.,][]{BeaumontRannala04, BonhommeEtAl08,MarjoramEtAl14,TechnowEtAl15}%
, molecular biology
\citep[e.g.,][]{PritchardEtAl99}%
, population biology
\citep
[e.g.,][]{EstoupEtAl04,NeuenschwanderEtAl08,BazinEtAl10,Beaumont10,BertorelleEtAl10,EstoupGuillemaud10,WegmannExcoffier10,AscunceEtAl11,DrovandiPettitt11b}%
, population genetics
\citep
[e.g.,][]{BeaumontEtAl02,MarjoramEtAl03,TallmonEtAl04,HamiltonEtAl05,TanakaEtAl06,BeaumontEtAl09,EstoupEtAl12}
and phylogeography
\citep[e.g.,][]{Beaumont10,HickersonEtAl06}%
, synthetic biology
\citep[e.g.,][]{BarnesEtAl11}%
, systematic biology
\citep[e.g.,][]{TavareEtAl97,TavareEtAl02,BaudetEtAl14,LintusaariEtAl17}%
, and systems biology
\citep
[e.g.,][]{Tavare05,RatmannEtAl07,RatmannEtAl09,ToniEtAl09,Beaumont10,ToniStumpf10,LiepeEtAl12,KoutroumpasEtAl16,LeniveEtAl16,LiepeStumpf18}%
; climate science
\citep[e.g.,][]{HoldenEtAl18}%
; economics including econometrics, finance, insurance
\citep
[e.g.,][]{PetersSisson06,PetersEtAl12,CalvetCzellar14,Picchini14,PetersEtAl18}%
; epidemiology
\citep
[e.g.,][]{SissonEtAl07,LucianiEtAl09,McKinleyEtAl09,ToniEtAl09,Beaumont10,BlumTran10,AandahlEtAl12,VolzEtAl12,MardulynEtAl13,KypraiosEtAl17,GroendykeWelch18,MckinleyEtAl18,RodriguesEtAl18}%
; hydrology
\citep[e.g.,][]{VrugtSadegh13,NottEtAl14}%
, nonlinear system identification (SID)
\citep[e.g.,][]{KrishnanathanEtAl16}%
; nuclear medicine
\citep[e.g.,][]{FanEtAl18}%
; oncology
\citep[e.g.,][]{SeigneurinEtAl11}%
; pharmacokinetics and pharmacodynamics
\citep[e.g.,][]{BechauxEtAl14,Picchini14}%
; psychology and psychometrics
\citep[e.g.,][]{TurnerVanZandt12,KangasraasioEtAl17,Karabatsos17c}%
; signal processing
\citep[e.g.,][]{PetersEtAl10}%
; stereology
\citep[e.g.,][]{BortotEtAl07}%
; and structural dynamics
\citep[e.g.,][]{AbdessalemEtAl18,ChiachioEtAl14}%
.

ABC methods have been developed to estimate various Bayesian models with
intractable likelihood, including alpha-stable
\citep[][]{PetersEtAl12}%
, bivariate beta distribution
\citep[][]{CrackelFlegal17}%
, coalescent
\citep[e.g.,][]{TavareEtAl97,FanKubatko11}%
, copula
\citep[][]{GrazianLiseo15,GrazianLiseo17semip,GrazianLiseo17chap}%
, differential equation
\citep[e.g.,][]{ToniEtAl09}%
, ecological
\citep[e.g.,][]{Wood10b, FasioloWood18}%
, epidemic
\citep[e.g.,][]{McKinleyEtAl09,McKinleyEtAl14,MckinleyEtAl18,KypraiosEtAl17}%
, extreme value
\citep[][]{ErhardtSmith12,ErhardtSisson16}%
, financial
\citep[e.g.,][]{PetersEtAl12}%
, hidden Markov
\citep[e.g.,][]{JasraEtAl12}%
, hydrological
\citep[e.g.,][]{NottEtAl14}%
, image analysis
\citep[e.g.,][]{NottEtAl14,MooresEtAl15}%
, network analysis
\citep[e.g.,][]{WangAtchade14,FayEtAl15}%
, order-restricted
\citep[][]{Karabatsos17c}%
, population evolution
\citep[e.g.,][]{MardulynEtAl13}%
, quantile distribution
\citep[e.g.,][]{AllinghamEtAl09,DrovandiPettitt11,Prangle17}%
, spatial process
\citep[][]{ShirotaGelfand16}%
, species abundance distribution
\citep[e.g.,][]{JabotEtAl11}%
, state-space
\citep[][]{BarthelmeChopin14,VakilzadehEtAl17}%
, stationary process
\citep[][]{AndradeRifo17}%
, statistical relational learning
\citep[][]{Cussens11}%
, susceptible-infected-removed (SIR)
\citep[][]{ToniEtAl09}%
, and time-series models
\citep[e.g.,][]{JasraEtAl14,Jasra15}%
. ABC\ methods have also been developed for optimal Bayesian designs
\citep[][]{DrovandiPettitt13,HainyEtAl16}%
, reinforcement learning
\citep[][]{DimitrakakisTziortziotis13}%
, and the estimation of intractable integrated likelihoods
\citep[][]{GrazianLiseo14}
and for approximate maximum likelihood estimation
\citep[e.g.,][]{RubioJohansen13, PicchiniAnderson17}%
; while a few studies of ABC asymptotics have emerged
\citep[e.g.,][]{DeanSingh11,BarberEtAl15,LiFearnhead16,LiFearnhead18}%
. Currently, there are at least 28 software packages and published program
code for ABC analysis
\citep
{BeaumontEtAl02,AndersonEtAl05,HickersonEtAl07,FollEtAl08,CornuetEtAl08,CornuetEtAl10,CornuetEtAl14,JobinMountain08,TallmonEtAl08,LopesEtAl09,Thornton09,BlumFrancois10,BrayEtAl10,LiepeEtAl10,PavlidisEtAl10,WegmannEtAl10,HuangEtAl11,KilbingerEtAl11,CsilleryEtAl12,JabotEtAl13,Picchini13,TsutayaYoneda13,BarthelmeChopin14,FasioloWood14,AkeretEtAl15,NunesPrangle15,Vrugt16,JenningsMadigan17,MertensEtAl18}%
. A recent Google Scholar search using the terms \textquotedblleft Approximate
Bayesian Computation\textquotedblright\ OR \textquotedblleft Synthetic
Likelihood\textquotedblright\ yielded 8,480 publications, and this number
continues to grow rapidly.

The vastity of the literature on ABC methods created a need to review and
relate all ABC approaches so that scientists can more readily understand and
apply them for their own work. This article provides a unifying review,
general representation, and classification of all ABC methods from the view of
approximate likelihood theory. Currently, any ABC method (algorithm)\ can be
categorized as either (1)\ rejection-, (2)\ kernel-, and (3)\ coupled ABC; and
(4)\ synthetic-, (5)\ empirical- and (6)\ bootstrap-likelihood methods; and
can be combined with classical MC or VI\ algorithms. However, given the vast
ABC literature, all of these methods may appear very different to scientists.
Further, all 22 reviews of ABC methods
\citep
{MarjoramEtAl03,PlagnolTavare04,MarjoramTavare06,Beaumont10,BertorelleEtAl10,CsilleryEtAl10,HartigEtAl11,Robert11,SissonFan11,MarinEtAl12,TurnerVanZandt12,BlumEtAl13,Marjoram13,SunnaakerEtAl13,BaragattiPudlo14,Robert16,ErhardtSisson16,KypraiosEtAl17,LintusaariEtAl17,Drovandi17,SissonEtAl18,DrovandiEtAl18}
have covered rejection and kernel ABC methods, but only three covered
synthetic likelihood, one reviewed the empirical likelihood, and none have
reviewed coupled ABC and bootstrap likelihood methods.

This article provides a unifying review, general representation, and
classifications of all ABC methods, from the approximate likelihood theory
view
\citep[][Ch.24]{EfronTibs93}%
. This clarifies how these methods can be characterized, related, combined,
improved, and applied for future research.

Next, Section 2 begins to set specific ideas by describing some examples of
Bayesian models defined by intractable likelihoods for which ABC is known to
be successful, while explaining ways in which likelihoods can be intractable.
Then, Section 3 presents, for all ABC\ methods, the general approximate
likelihood representation to unify and classify them, and a general iterative
IS (and MH)\ MC\ algorithm for sampling approximate posterior distributions.
On this basis, Section 4 reviews the six types of ABC methods; Section 5
itemizes ABC methods that have been combined with either the classical MC, VI,
or simulated annealing algorithms; Section 6 covers ABC model choice methods;
and Section 7 summarizes the available statistical software packages for ABC.
Section 8 describes some open problems in ABC. Section 9 concludes the article.

\section{Examples of Bayesian Models with Intractable Likelihoods}

A Bayesian model is defined by a parameter vector $\boldsymbol{\theta}$,
having space $\Theta\subseteq%
\mathbb{R}
^{d}$, a likelihood measure\ $f(\mathbf{y}_{n}\,|\,\boldsymbol{\theta})$ for a
given data set ($\mathbf{y}_{n}=(y_{i})_{i=1}^{n}$) of $n$ sample observations
with sample space $\mathcal{Y}_{n}$, and by a prior distribution
(measure)\ with density $\pi(\boldsymbol{\theta})$ defined on the parameter
space, $\Theta\subseteq%
\mathbb{R}
^{d}$. The likelihood and prior densities are each a continuous p.d.f. and/or
a discrete p.m.f., corresponding to c.d.f.s\ $F(\mathbf{y}_{n}%
\,|\,\boldsymbol{\theta})$ and $\Pi(\boldsymbol{\theta})$, respectively.

According to Bayes theorem, a set of data ($\mathbf{y}_{n}$) updates the prior
to a posterior distribution, defined by probability measure $\pi
(\boldsymbol{\theta}\,|\,\mathbf{y}_{n})=f(\mathbf{y}_{n}%
\,|\,\boldsymbol{\theta})\pi(\boldsymbol{\theta})/m(\mathbf{y}_{n})$, with
marginal likelihood normalizing constant $m(\mathbf{y}_{n})=%
{\textstyle\int}
f(\mathbf{y}_{n}\,|\,\boldsymbol{\theta})\Pi(\mathrm{d}\boldsymbol{\theta})$.
Also, $f_{n}(y)=%
{\textstyle\int}
f(y\,|\,\boldsymbol{\theta})\Pi($\textrm{d}$\boldsymbol{\theta}\,|\,\mathbf{y}%
_{n})$ is the posterior predictive density of a future observable $y$.
However, for many Bayesian models, the likelihood $f(\mathbf{y}_{n}%
\,|\,\boldsymbol{\theta})$ is analytically and/or computationally intractable.
Then, posterior inferences are infeasible using analytical, MCMC, SMC, PMC,
IS, VI, or other appropriate methods.

We now proceed to describe three examples of Bayesian models defined by
intractable likelihoods, each of which has a posterior distributions that can
be estimated using ABC methods.

\subsection{$g$-and-$k$ Distribution}

The first example of an intractable likelihood is defined by the $g$-and-$k$
distribution
\citep{MacGillivray92}%
.\ This distribution extends the normal distribution by allowing for added
skewness or heavier (lighter) tails, and can describe a wide variety
distribution shapes with four interpretable parameters. The generalized
$g$-and-$k$ distribution
\citep{MacGillivray92}
is defined by the quantile function:
\begin{equation}
F_{gk}^{-1}(u;A,B,g,k)=A+B(1+c\tanh[(g/2)z_{u}])z_{u}(1+z_{u}^{2}%
)^{k},\label{gk}%
\end{equation}
where $z_{u}=\mathrm{N}^{-1}(u\boldsymbol{\mid}0,1)$ is the standard normal
$\mathrm{N}(0,1)$ quantile function, and has parameters $\boldsymbol{\theta
}=(A,B,g,k)$. Here, $A\in%
\mathbb{R}
$ is a location parameter, $B>0$ is a scale parameter, $g\geq0$ controls
skewness, $k\geq0$ controls kurtosis (tail size), and $c=.8$ provides a
standard choice of overall asymmetry constant
\citep{RaynerMacGillivray02,MacGillivray86}%
.

The likelihood p.d.f. of the $g$-and-$k$ distribution has the form
$f(\mathbf{y}_{n}\,|\,\boldsymbol{\theta})=%
{\textstyle\prod\nolimits_{i=1}^{n}}
f(y_{i}\,|\,\boldsymbol{\theta})$, assuming that the given set of $n$ sample
observations $\mathbf{y}_{n}=(y_{i})_{i=1}^{n}$ are i.i.d. from
$f(y\,|\,\boldsymbol{\theta})$. A Bayesian $g$-and-$k$ model is completed by
the specification of a prior distribution $\pi(\boldsymbol{\theta})$ on the
space of the model parameters $\boldsymbol{\theta}=(A,B,g,k)$. However, the
$g$-and-$k$ likelihood p.d.f. has no closed-form expression in general.
Instead, the likelihood p.d.f. is expressible in terms of derivatives of
quantile functions, and needs to be computed completely numerically for each
of the individual data points $y_{i}$
\citep{RaynerMacGillivray02,Prangle17}%
. As a result, the computation of the $g$-and-$k$ likelihood is slow even for
a moderate data sample size $n$, and hundreds of time slower than computing
the normal p.d.f.
\citep{RaynerMacGillivray02,Prangle17}%
. Then, for the inference from the posterior distribution $\pi
(\boldsymbol{\theta}\,|\,\mathbf{y}_{n})\propto f(\mathbf{y}_{n}%
\,|\,\boldsymbol{\theta})\pi(\boldsymbol{\theta})$ of the Bayesian $g$-and-$k$
model, any standard MCMC\ approach is computationally slow because it requires
making $n$ calls to numerical optimization to evaluate the likelihood p.d.f.
$f(\mathbf{y}_{n}\,|\,\boldsymbol{\theta})$ in each MCMC\ sampling iteration
\citep{Prangle17}%
.

ABC can be employed for performing inferences from the approximate posterior
distribution of the parameters $\boldsymbol{\theta}=(A,B,g,k)$, based on the
specification of a surrogate likelihood that approximates the exact
$g$-and-$k$ model likelihood, but is less computationally costly. The
rejection ABC\ (R-ABC) method (Section 4.1) has proven to be a viable
ABC\ method for this model, which is based on finding parameter values which
produce simulated (synthetic)\ data sets that are similar to the observed data
sets $\mathbf{y}_{n}$, based on summary statistics of each data set
\citep[for details, see][]{Prangle17}%
.

\subsection{Mixed-Effects Model}

The second example is given by the general mixed-effect model, which for the
$j$th observation within blocking factor $k$, is given by:%
\begin{equation}
y_{jk}=\mathbf{x}_{jk}^{\intercal}\boldsymbol{\beta}+b_{k}+\varepsilon
_{jk},\label{mixEff}%
\end{equation}
for $j=1,\ldots,J_{k}$ and $k=1,\ldots,K$, and $n=%
{\textstyle\sum\nolimits_{k=1}^{K}}
J_{k}$, where the $\mathbf{x}_{jk}$ are the fixed-effect predictor vectors,
assuming $b_{k}\overset{\text{iid}}{\sim}F_{\boldsymbol{\zeta}}$ and
$\varepsilon_{jk}\overset{\text{iid}}{\sim}F_{\boldsymbol{\sigma}}$ with the
$b_{k}$ and the $\varepsilon_{jk}$ uncorrelated. The mixed-effects model
(\ref{mixEff}) has corresponding likelihood:%
\begin{equation}
f(\mathbf{y}_{n}\,|\,\boldsymbol{\theta})=%
{\displaystyle\int}
{\textstyle\prod\limits_{j,k}}
G(y_{jk}-\mathbf{x}_{jk}^{\intercal}\boldsymbol{\beta}-b_{k})\mathrm{d}%
{\textstyle\prod\limits_{k}}
F_{b}(b_{k}).\label{mixedLike}%
\end{equation}
A Bayesian mixed-effects model is completed by the specification of a prior
distribution $\pi(\boldsymbol{\theta})=\pi(\boldsymbol{\beta}%
,\boldsymbol{\zeta},\boldsymbol{\sigma})$, with corresponding posterior p.d.f.
$\pi(\boldsymbol{\theta}\,|\,\mathbf{y}_{n})\propto f(\mathbf{y}%
_{n}\,|\,\boldsymbol{\theta})\pi(\boldsymbol{\theta})$.

The likelihood (\ref{mixedLike}) is intractable virtually impossible to
evaluate unless $F_{\boldsymbol{\zeta}}$ and $F_{\boldsymbol{\sigma}}$ are
standard distributions such as normal distributions. But it is well-known that
empirical violations of these assumptions would cast doubt about the accuracy
of the inferences from the posterior $\pi(\boldsymbol{\theta}\,|\,\mathbf{y}%
_{n})$. In principle, any of the Approximate Bayesian Computational (ABC)
methods (Section 4) can be implemented to perform posterior inferences of the
Bayesian mixed-effects model with intractable likelihood.

\subsection{Hidden Potts Model}

The third example is provided by the Potts model, which originated in
statistical physics, and is now widely used for applications in image
processing, spatial modelling, computational biology, and computational
neuroscience. To explain this model, let $i\in\{1,\ldots,n\}$ denote the
pixels (nodes)\ of an image lattice, where for each node $i$, the value
$y_{i}\in\{1,\ldots,k\}$ denotes the node's state among $k$ possible states.
The Potts model is a Markov random field model defined in terms of its
conditional probabilities:%
\begin{equation}
\Pr(y_{i}\,|\,y_{i\sim\ell})=\dfrac{\exp\{\theta%
{\textstyle\sum\nolimits_{i\sim\ell}}
\delta(y_{i},y_{\ell})\}}{%
{\textstyle\sum\nolimits_{j=1}^{k}}
\{\theta%
{\textstyle\sum\nolimits_{i\sim\ell}}
\delta(y_{i},y_{\ell})\}},
\end{equation}
for $i=1,\ldots,n$, where $\theta\geq0$ is the inverse temperature (scale)
parameter, $i\sim\ell$ are the neighboring pixels of $i$, and $\delta
(\cdot,\cdot)$ is the Kronecker delta function. For example, using a
first-order neighborhood, $i\sim\ell$ refer to the four pixels immediately
adjacent to internal node $i$ of the image lattice, and pixels on the image
boundary have less than four neighbors.

A Bayesian Potts model is completed by the specification of a prior
distribution $\pi(\theta)$ on $[0,\infty)$. According to Bayes' theorem, given
the data $\mathbf{y}_{n}=(y_{i})_{i=1}^{n}$, the posterior distribution of the
Potts model is given by $\pi(\theta\,|\,\mathbf{y}_{n})\propto f(\mathbf{y}%
_{n}\,|\,\theta)\pi(\theta)$, with likelihood defined by:%
\begin{equation}
f(\mathbf{y}_{n}\,|\,\boldsymbol{\theta})=\dfrac{\exp\left(  \theta%
{\textstyle\sum\limits_{i\sim\ell\in\mathcal{E}}}
\delta(y_{i},y_{\ell})\right)  }{%
{\textstyle\sum\limits_{\mathbf{y}_{n}^{\ast}\in\mathcal{\Im}}}
\exp\left(  \theta%
{\textstyle\sum\limits_{i\sim\ell\in\mathcal{E}}}
\delta(y_{i}^{\ast},y_{\ell}^{\ast})\right)  }.\label{Pottslike}%
\end{equation}
The Potts model likelihood (\ref{Pottslike}) is intractable because its
denominator involves a sum over all $k^{n}$ possible combinations of the
labels $\mathbf{y}_{n}\in\mathcal{\Im}$. Clearly, the likelihood computation
time, and the inference of the posterior distribution $\pi(\theta
\,|\,\mathbf{y}_{n})$, depends on the size $n$ of the image lattice.

For the inference of the posterior distribution $\pi(\theta\,|\,\mathbf{y}%
_{n})$, for large $n$, an MCMC algorithm would be virtually impossible to
implement because it would require evaluating the likelihood, and rejection
ABC\ (R-ABC) is time consuming because simulating a (synthetic)\ data set from
the likelihood (\ref{Pottslike}) is computationally costly. However, the
synthetic likelihood (SL-ABC) and bootstrap likelihood (BL-ABC) approaches to
ABC have proven to be successful in providing inference from the posterior
distribution $\pi(\theta\,|\,\mathbf{y}_{n})$ of the Bayesian Potts model with
relatively low computational cost
\citep[see][]{MooresEtAl15,ZhuMarinLeisen16}%
. the general SL-ABC and BL-ABC methods are described in Section 4.

\section{Approximate Likelihoods and Sampling Algorithm}

We now introduce the general unifying representation of all ABC methods. Each
ABC method provides approximate inference of the posterior distribution
$\pi(\boldsymbol{\theta}\,|\,\mathbf{y}_{n})=f(\mathbf{y}_{n}%
\,|\,\boldsymbol{\theta})\pi(\boldsymbol{\theta})/m(\mathbf{y}_{n})$ for a
given Bayesian model defined by an intractable likelihood $f(\mathbf{y}%
_{n}\,|\,\boldsymbol{\theta})$, and by implication, an intractable marginal
likelihood $m(\mathbf{y}_{n})$.

Specifically, ABC provides tractable posterior inference by replacing the
exact likelihood with an approximate likelihood that admits the general
representation:%
\begin{equation}
L_{\boldsymbol{\eta}}(\mathbf{y}_{n}\,|\,\boldsymbol{\theta})=%
{\displaystyle\int}
K(\mathbf{t}(\mathbf{y}_{n})\,|\,\widehat{\boldsymbol{\eta}}%
_{\boldsymbol{\theta},N}^{\{\mathbf{t}(\mathbf{z}_{n(k)})\}})%
{\textstyle\prod\nolimits_{k=1}^{N}}
f(\mathbf{z}_{n(k)}\,|\,\boldsymbol{\theta})\mathrm{d}\mathbf{z}_{n(k)}.
\label{Surrogate}%
\end{equation}
Above, $\mathbf{t}(\mathbf{y}_{n})$ is a vector of summary statistics of the
data $\mathbf{y}_{n}$ (possibly, $\mathbf{t}(\mathbf{y}_{n})\equiv
\mathbf{y}_{n}$), which ideally are of low dimension (\textrm{dim}) and
sufficient as the given Bayesian model and data set permit; parameter
identifiability requires $\dim(\mathbf{t})\geq\dim(\boldsymbol{\theta})$.
$K(\mathbf{t}(\mathbf{y}_{n})\,|\,\widehat{\boldsymbol{\eta}}%
_{\boldsymbol{\theta},N}^{\{\mathbf{t}(\mathbf{z}_{n(k)})\}})$ is the kernel
density function, given a parameter estimate $\widehat{\boldsymbol{\eta}%
}_{\boldsymbol{\theta},N}^{\{t(\mathbf{z}_{n(k)})\}}$ obtained from
$N$\ synthetic data set samples $\{\mathbf{z}_{n(k)}\}_{k=1}^{N}%
\overset{\text{iid}}{\sim}f(\mathbf{\cdot}\,|\,\boldsymbol{\theta})$ of size
$n(k)$ from the exact model likelihood. ABC\ methods that use $N=1$ set
$\mathbf{t}(\mathbf{z}_{n(k)})\equiv\mathbf{t}(\mathbf{z}_{n})$, and methods
that sample no synthetic data ($N=0$) set $K(\mathbf{t}(\mathbf{y}%
_{n})\,|\,\widehat{\boldsymbol{\eta}}_{\boldsymbol{\theta},N}^{\{t(\mathbf{z}%
_{n(k)})\}})\equiv K(\mathbf{t}(\mathbf{y}_{n})\,|\,\widehat{\boldsymbol{\eta
}}_{\boldsymbol{\theta},N})$.

An unbiased estimator of the general approximate likelihood (\ref{Surrogate})
is given by:%
\begin{equation}
\widehat{L}_{\boldsymbol{\eta}}^{(S)}(\mathbf{y}_{n}\,|\,\boldsymbol{\theta
})=\dfrac{1}{S}%
{\displaystyle\sum\limits_{s=1}^{S}}
K(\mathbf{t}(\mathbf{y}_{n})\,|\,\widehat{\boldsymbol{\eta}}%
_{\boldsymbol{\theta},N}^{\{t(\mathbf{z}_{n(k)})\}}),\qquad\{\mathbf{z}%
_{n(k)}\}_{k=1}^{N}\overset{\text{iid}}{\sim}f(\mathbf{\cdot}%
\,|\,\boldsymbol{\theta}). \label{surLikeEst}%
\end{equation}
For discrete data, the indicator function kernel $K(\mathbf{t}(\mathbf{y}%
_{n})\,|\,\widehat{\boldsymbol{\eta}}_{\boldsymbol{\theta},N}^{\{t(\mathbf{z}%
_{n(k)})\}})=\mathbf{1}(\mathbf{t}(\mathbf{y}_{n})=\mathbf{t}(\mathbf{z}%
_{n}))$ provides an unbiased estimator of the exact model likelihood
$f(\mathbf{y}_{n}\,|\,\boldsymbol{\theta})$
\citep[][]{Rubin84}%
. For realistic settings, the equality constraint is replaced by some kernel
$K$\ in (\ref{surLikeEst}) that tolerates some small level of inequality.

The general likelihood (\ref{Surrogate}) gives rise to the approximate
posterior density:
\begin{equation}
\pi_{L}(\boldsymbol{\theta}\,|\,\mathbf{y}_{n})=L_{\boldsymbol{\eta}%
}(\mathbf{y}_{n}\,|\,\boldsymbol{\theta})\pi(\boldsymbol{\theta}%
)/m_{L}(\boldsymbol{\theta}), \label{ProxPost}%
\end{equation}
with $m_{L}(\boldsymbol{\theta})=%
{\textstyle\int}
L_{\boldsymbol{\eta}}(\mathbf{y}_{n}\,|\,\boldsymbol{\theta})\Pi
(\mathrm{d}\boldsymbol{\theta})$.\ Theoretically, (\ref{ProxPost}) exemplifies
a \textit{limited information likelihood} posterior when $\boldsymbol{\eta
}_{\boldsymbol{\theta}}\equiv\boldsymbol{\theta}$ and $\mathbf{t}%
(\mathbf{y}_{n}):\neq\mathbf{y}_{n}$
\citep[][]{PrattRaiffaSchlaifer65,DoksumLo90,Zellner97,Kwan99,Kim02}%
, and \textit{indirect inference}
\citep[][]{GourierouxEtAl93,HegglandFrigessi04,JiangTurnbull04}
when $\boldsymbol{\eta}_{\boldsymbol{\theta}}\neq\boldsymbol{\theta}$
\citep[][]{DrovandiEtAl11}%
.

Using the general approximate likelihood representation (\ref{Surrogate}),
Table 1 summarizes the six types of ABC methods proposed in the literature.
They differ by the choice of kernel function ($K$), the number ($N$) and
method of sampling iid synthetic data sets from the exact model likelihood
$f(\mathbf{y}_{n}\,|\,\boldsymbol{\theta})$, and by whether or not it
implements indirect inference.

For any one of these six types of ABC methods, a general IS algorithm can be
employed for ABC\ inference of the posterior $\pi_{L}(\boldsymbol{\theta
}\,|\,\mathbf{y}_{n})$ in (\ref{ProxPost}), for any function
$g(\boldsymbol{\theta})$ of interest, using the prior $\pi(\boldsymbol{\theta
})$ as the instrumental density. For the general inference of $%
{\textstyle\int}
g(\boldsymbol{\theta})\pi_{L}(\boldsymbol{\theta}\,|\,\mathbf{y}%
_{n})\mathrm{d}\boldsymbol{\theta}$, this integral can be rewritten as:%
\begin{equation}
\frac{%
{\textstyle\int}
g(\boldsymbol{\theta})L_{\boldsymbol{\eta}}(\mathbf{y}_{n}%
\,|\,\boldsymbol{\theta})\Pi(\mathrm{d}\boldsymbol{\theta})}{%
{\textstyle\int}
L_{\boldsymbol{\eta}}(\mathbf{y}_{n}\,|\,\boldsymbol{\theta})\Pi
(\mathrm{d}\boldsymbol{\theta})},
\end{equation}
and estimated via IS\ by employing a sampling scheme of the form:
\begin{subequations}
\begin{equation}
\frac{\tfrac{1}{S}%
{\textstyle\sum\nolimits_{s=1}^{S}}
g(\boldsymbol{\theta}_{s})K(\mathbf{t}(\mathbf{y}_{n}%
)\,|\,\widehat{\boldsymbol{\eta}}_{\boldsymbol{\theta}_{s},N}^{\{\mathbf{t}%
(\mathbf{z}_{n(k)}^{(s)})\}})}{\tfrac{1}{S}%
{\textstyle\sum\nolimits_{s=1}^{S}}
K(\mathbf{t}(\mathbf{y}_{n})\,|\,\widehat{\boldsymbol{\eta}}%
_{\boldsymbol{\theta}_{s},N}^{\{\mathbf{t}(\mathbf{z}_{n(k)}^{(s)})\}}%
)},\text{ }%
\end{equation}
with $\{\mathbf{z}_{n(k)}^{(s)}\}_{k=1}^{N}|\boldsymbol{\theta}_{s}\sim
f(\mathbf{\cdot}\,|\,\boldsymbol{\theta}_{s}),$ and $\{\boldsymbol{\theta}%
_{s}\}_{s=1}^{S}\overset{\text{iid}}{\sim}\pi(\boldsymbol{\theta})$.

\noindent The general IS algorithm is given by:

\bigskip
\end{subequations}
\begin{description}
\item[\textbf{ABC}\ \textbf{Importance Sampling Algorithm (ABC-IS)}]
\textit{\newline}\noindent for $s=1$ to $S$ do\newline\noindent$%
\begin{array}
[c]{l}%
(a)\ \text{Sample from prior, }\boldsymbol{\theta}_{s}\sim\pi
(\boldsymbol{\theta})\text{;}\newline\\
(b)\ \text{Find estimate }\widehat{\boldsymbol{\eta}}_{\boldsymbol{\theta}%
_{s},N}^{\{\mathbf{t}(\mathbf{z}_{n(k)}^{(s)})\}}\text{ from }N\text{ sampled
data sets of size }n(k),\\
\ \ \ \ \ \ \{\mathbf{z}_{n(k)}^{(s)}=(z_{i}^{(s)})_{i=1}^{n(k)}\}_{k=1}%
^{N}\overset{\text{iid}}{\sim}f(\mathbf{y}_{n}\,|\,\boldsymbol{\theta}%
_{s})\text{ \ (if ABC method uses }N\geq1\text{); }\\
(c)\ \text{Set the IS\ weight for }\boldsymbol{\theta}_{s}\text{ to }%
\omega_{s}\equiv K(\mathbf{t}(\mathbf{y}_{n})\,|\,\widehat{\boldsymbol{\eta}%
}_{\boldsymbol{\theta}_{s},N}^{\{\mathbf{t}(\mathbf{z}_{n(k)}^{(s)}%
)\}});\text{ }\newline%
\end{array}
$\newline end for.\newline
\end{description}

\noindent The ABC-IS algorithm yields output of $S$ samples of the model
parameters and normalized sampling weights, $(\boldsymbol{\theta}%
_{s},\overline{\omega}_{s}=\omega_{s}/%
{\textstyle\sum\nolimits_{j=1}^{S}}
\omega_{j})_{s=1}^{S}$, which can be used to construct estimates of the
posterior $\pi_{L}(\boldsymbol{\theta}\,|\,\mathbf{y})$ for parameter
functions of interest. The convergence of the output can be evaluated by the
Effective Sample Size statistic, $\mathrm{ESS}=1\left/
{\textstyle\sum\nolimits_{s=1}^{S}}
\{\overline{\omega}_{s}/%
{\textstyle\sum\nolimits_{s=1}^{S}}
\overline{\omega}_{s}\}^{2}\right.  ,$ which ranges from $1$ to $S$ (perfect
outcome where $(\boldsymbol{\theta}_{s})_{s=1}^{S}$ are iid)
\citep[][]{Liu01}%
.

The ABC-IS algorithm is conceptually simple, and can be easily parallelized
because Step $(a)$\ draws independent samples from the prior $\pi
(\boldsymbol{\theta})$ over iterations. Also, this algorithm can be applied to
any one of the six types of ABC methods, with respect to specific choices of
$N$ (for Step $(b)$) and the kernel function $K$ defining the importance
sampling weight (for Step $(c)$), as summarized in second and third columns of
Table 1. More details in Section 4. We want to stress that some users of the 6
different ABC methods may not at first immediately recognize the ABC-IS
algorithm, but the importance sampling feature is indeed an integral part of
these methods. For example, the popular, rejection ABC (R-ABC) method employs
the same algorithmic Step $(a)$ to generate a prior sample $\boldsymbol{\theta
}_{s}$; and then samples $N=1$ synthetic data set $\mathbf{z}_{n(k)}^{(s)}$ in
Step $(b)$ from the model likelihood $f(\mathbf{y}_{n}\,|\,\boldsymbol{\theta
}_{s})$; and then in Step $(c)$ employs an IS\ weight defined by a binary (0
or 1) valued function indicating whether the summarized data $\mathbf{t}%
(\mathbf{y}_{n})$ is close in distance to the summary $\mathbf{t}%
(\mathbf{z}_{n(k)}^{(s)})$ of the sampled synthetic data set.

Alternatively, an MH algorithm can be used (with weights $\omega_{s}\equiv1$),
by changing the ABC-IS\ algorithm so that step (a)\ draws $\boldsymbol{\theta
}^{\ast}\sim q(\boldsymbol{\theta}_{s}\,|\,\boldsymbol{\theta}_{s-1})$ from a
proposal density (distribution) $q$, and step $(c)$\ accepts
$\boldsymbol{\theta}_{s}\equiv\boldsymbol{\theta}^{\ast}$ with probability:
\begin{equation}
\min\left\{  1,\dfrac{K(\mathbf{t}(\mathbf{y}_{n}%
)\,|\,\widehat{\boldsymbol{\eta}}_{\boldsymbol{\theta}^{\ast},N}%
^{\{\mathbf{t}(\mathbf{z}_{n(k)s})\}})\pi(\boldsymbol{\theta}^{\ast
})q(\boldsymbol{\theta}_{s-1}\,|\,\boldsymbol{\theta}_{s})}{K(\mathbf{t}%
(\mathbf{y}_{n})\,|\,\widehat{\boldsymbol{\eta}}_{\boldsymbol{\theta}_{s-1}%
,N}^{\{\mathbf{t}(\mathbf{z}_{n(k)s-1})\}})\pi(\boldsymbol{\theta}%
_{s-1})q(\boldsymbol{\theta}^{\ast}\,|\,\boldsymbol{\theta}_{s-1})}\right\}  ,
\end{equation}
based on the MH\ acceptance ratio.

Either the ABC-IS algorithm or the Metropolis algorithm can be extended to
versions (resp.) which adaptively find the optimal instrumental or proposal
density (resp.) over iterations, with the aim of yielding Monte Carlo samples
of $\boldsymbol{\theta}$ that more quickly converge to samples from the
posterior distribution $\pi_{L}(\boldsymbol{\theta}\,|\,\mathbf{y}_{n})$. This
includes Population Monte Carlo (PMC)
\citep{CappeEtAl04}%
, adaptive multiple importance sampling
\citep{CornuetEtAl12}%
, and adaptive Metropolis algorithms
\citep[e.g.,][]{RobertsRosenthal09}%
, among others.

The following review of ABC\ methods is cast within the ABC-IS\ algorithm, for
simplicity and without loss of generality, unless otherwise indicated.%

\begin{table}[tbp] \centering
\begin{tabular}
[c]{|c|c|c|c|}\hline\hline
\multicolumn{1}{||c|}{$%
\begin{array}
[c]{c}%
\text{ }\\
\text{ }\\
\text{ABC}\\
\text{Method}\\
\text{ }\\
\text{ }%
\end{array}
$} & $%
\begin{array}
[c]{l}%
\text{ }\\
\text{Kernel\textbf{, }}K\text{, of approximate likelihood:}\\
L_{\boldsymbol{\eta}}(\mathbf{y}_{n}\,|\,\boldsymbol{\theta})=\\%
{\displaystyle\int}
K(\mathbf{t}(\mathbf{y}_{n})\,|\,\widehat{\boldsymbol{\eta}}%
_{\boldsymbol{\theta},N}^{\{\mathbf{t}(\mathbf{z}_{n(k)})\}})%
{\textstyle\prod\limits_{k=1}^{N}}
f(\mathbf{z}_{n(k)}\,|\,\boldsymbol{\theta})\mathrm{d}\mathbf{z}_{n(k)}\\
\text{ }%
\end{array}
$ & $N$ & \multicolumn{1}{|c||}{$%
\begin{array}
[c]{c}%
\text{Indirect}\\
\text{Inference}%
\end{array}
$}\\\hline\hline
$%
\begin{array}
[c]{c}%
\text{ }\\
\text{Rejection}\\
\text{ }%
\end{array}
$ & $\mathbf{1}(\rho_{\mathbf{t}(\mathbf{y}_{n}),\mathbf{t}(\mathbf{z}_{n}%
)}\leq\epsilon)$ & $1$ & $\boldsymbol{\eta}\equiv\rho$\\\hline
$%
\begin{array}
[c]{c}%
\text{ }\\
\text{Kernel}\\
\text{ }%
\end{array}
$ & $K(\rho_{\mathbf{t}(\mathbf{y}_{n}),\mathbf{t}(\mathbf{z}_{n})})\text{,
smooth}$ & $1$ & $\boldsymbol{\eta}\equiv\rho$\\\hline
$%
\begin{array}
[c]{c}%
\text{ }\\
\text{Coupled}\\
\\
\text{ }%
\end{array}
$ & $%
\begin{array}
[c]{c}%
\text{ }\\%
{\textstyle\int}
\mathbf{1}(\rho_{\mathbf{t}(\mathbf{y}_{n}),\mathbf{t}(\mathbf{z}%
_{n}\{\mathbf{u},\boldsymbol{\theta}\})})\pi(\mathbf{u})\mathrm{d}\mathbf{u}\\
\pi(\mathbf{u}):[0,1]^{q}\rightarrow\lbrack0,\infty)\\
\text{ }%
\end{array}
$ & $0$ & $\boldsymbol{\eta}\equiv\rho$\\\hline
$%
\begin{array}
[c]{c}%
\text{ }\\
\text{SL:\ }\\
\text{Synthetic}\\
\text{Likelihood}\\
\text{ }%
\end{array}
$ & $%
\begin{array}
[c]{c}%
\mathrm{n}(\mathbf{t}(\mathbf{y}_{n}\mathbf{)}\,|\,\widehat{\boldsymbol{\mu}%
}_{\boldsymbol{\theta},N}^{\{\mathbf{t}(\mathbf{z}_{n(k)})\}}%
,\widehat{\mathbf{\Sigma}}_{\boldsymbol{\theta},N}^{\{\mathbf{t}%
(\mathbf{z}_{n(k)})\}})\\
\text{or }K(\mathbf{t}(\mathbf{y}_{n})\,|\,\widehat{\boldsymbol{\eta}%
}_{\boldsymbol{\theta},N}^{\{t(\mathbf{z}_{n(k)})\}})
\end{array}
$ & $%
\genfrac{}{}{0pt}{0}{\text{$>$$>$}1}{\text{{\small large}}}%
$ & $%
\genfrac{}{}{0pt}{0}{\boldsymbol{\eta}\equiv(\boldsymbol{\mu},\mathbf{\Sigma
})}{\text{or }\boldsymbol{\eta}\equiv K}%
$\\\hline
$%
\begin{array}
[c]{c}%
\text{ }\\
\text{EL:\ }\\
\text{Empirical }\\
\text{Likelihood}\\
\text{ }%
\end{array}
$ & $%
\begin{array}
[c]{c}%
\widehat{L}_{n}^{el}(\mathbf{y}_{n}\,|\,\boldsymbol{\theta})=%
\begin{array}
[c]{c}%
\max\\%
\genfrac{}{}{0pt}{1}{\{\mathbf{p}(\boldsymbol{\theta})\,:\mathbf{\,p}%
\in\lbrack0,1]^{n};}{\mathbb{E}_{F}[h(Y,\boldsymbol{\theta})]=0\}}%
\end{array}%
{\textstyle\prod\limits_{i=1}^{n}}
p_{i}(\boldsymbol{\theta})\\
p_{i}\in\lbrack0,1],\text{ \ }%
{\textstyle\sum\nolimits_{i=1}^{n}}
p_{i}=1
\end{array}
$ & $0$ & $\boldsymbol{\eta}\equiv\mathbf{p}(\boldsymbol{\theta})$\\\hline
$%
\begin{array}
[c]{c}%
\text{ }\\
\text{BL:}\\
\text{Bootstrap }\\
\text{Likelihood}\\
\text{ }\\
\text{ }%
\end{array}
$ & $%
\begin{array}
[c]{c}%
\text{ }\\
\widehat{f}_{n}(\mathbf{y}_{n}\,|\,\boldsymbol{\theta})\text{ (or
}\{\widehat{f}_{n,s}(\mathbf{y}_{n}\,|\,\boldsymbol{\theta})\}_{s=1}%
^{S}\text{)}\\
\text{estimated from nested, single,}\\
\text{ }\tbinom{n}{m}\text{{\small , }or }n\text{/}m\text{ bootstrap of data
}\mathbf{y}_{n}\text{ }\\
\text{(per MC iteration }s\text{)}\\
\text{ }%
\end{array}
$ & $0$ & No\\\hline
\multicolumn{4}{|l|}{}\\
\multicolumn{4}{|l|}{$\mathbf{y}_{n}\text{: data set, size }n\text{; }$
$f(\cdot\,\,|\,\boldsymbol{\theta})\text{: original/exact\ model likelihood;}%
$}\\
\multicolumn{4}{|l|}{$\{\mathbf{z}_{n(k)}\}_{k=1}^{N}$ are $N\text{ samples of
synthetic data sets }\mathbf{z}_{n(k)}\text{ of size }n(k);$}\\
\multicolumn{4}{|l|}{$\mathbf{t}(\cdot)\ $data $\text{summary statistic
(vector); }\mathbf{1}(\cdot)$: indicator function;}\\
\multicolumn{4}{|l|}{$\rho_{\mathbf{t}(\mathbf{y}_{n}),\mathbf{t}%
(\mathbf{z}_{n})}\text{ (e.g. Euclidean) distance of }\mathbf{t}%
(\mathbf{z}_{n})\text{ to }\mathbf{t}(\mathbf{y}_{n})\text{; }\epsilon>0\text{
tolerance;}$}\\
\multicolumn{4}{|l|}{$K(\mathbf{t}(\mathbf{y}_{n}%
)\,|\,\widehat{\boldsymbol{\eta}}_{\boldsymbol{\theta},N}^{\{\mathbf{t}%
(\mathbf{z}_{n(k)})\}})\text{: density of }\mathbf{t}(\mathbf{y}_{n})\text{,
given estimate }\widehat{\boldsymbol{\eta}}\text{ from }\{\mathbf{t}%
(\mathbf{z}_{n(k)})\}_{k=1}^{N};$}\\
\multicolumn{4}{|l|}{$\mathrm{n}(\mathbf{t}(\mathbf{y}_{n}\mathbf{)}%
\,|\,\widehat{\boldsymbol{\mu}},\widehat{\mathbf{\Sigma}})\text{: multivar.
normal p.d.f., given estimate }\widehat{\boldsymbol{\eta}}%
=(\widehat{\boldsymbol{\mu}},\widehat{\mathbf{\Sigma}})\text{;}$}\\
\multicolumn{4}{|l|}{$\mathbb{E}_{f}[h(Y,\boldsymbol{\theta})]\text{: function
}h(Y,\boldsymbol{\theta})\text{ expectation over }f(y\,\,|\,\boldsymbol{\theta
})\text{; }$}\\
\multicolumn{4}{|l|}{$\widehat{f}_{n}\text{: estimate of }f(\cdot
\,\,|\,\boldsymbol{\theta});$}\\
\multicolumn{4}{|c|}{}\\\hline
\multicolumn{4}{c}{}%
\end{tabular}
\caption{The six types of ABC methods, by kernel (K), and number (N) of synthetic data sets sampled per IS iteration.}%
\end{table}%

\section{ABC\ Methods}

Sections 4.1-4.6 review the six types of ABC\ methods (Table 1). They each can
be combined with other MC, VI, or simulated annealing algorithms, as mentioned
in Section 5.

\subsection{Rejection ABC (R-ABC)}

The R-ABC method is the oldest
\citep{FuLi97,WeissVonHaeseler98,PritchardEtAl99,BeaumontEtAl02}%
. R-ABC draws $N=1$ synthetic data set $\mathbf{z}_{n}\sim f(\mathbf{\cdot
}\,|\,\boldsymbol{\theta})$ in ABC-IS\ step (b), and then in step $(c)$, it
accepts the sample $\boldsymbol{\theta}_{s}$ when the distance $\rho
_{\mathbf{t}(\mathbf{y}_{n}),\mathbf{t}(\mathbf{z}_{n})}$ between
$\mathbf{t}(\mathbf{y}_{n})$ and $\mathbf{t}(\mathbf{z}_{n})$ is within a
chosen tolerance $\epsilon\geq0$, thereby using the IS\ weight $\omega
_{s}\equiv\mathbf{1}(\rho_{\mathbf{t}(\mathbf{y}_{n}),\mathbf{t}%
(\mathbf{z}_{n})}\leq\epsilon)$ with indicator (kernel)\ function
$\mathbf{1}(\cdot)$. The sampling algorithm is run until the desired number of
acceptances is obtained. Actually, the first R-ABC approach
\citep{TavareEtAl97}
did not employ a sampling step, but instead based rejection decisions on a
maximum likelihood estimate $\widehat{\boldsymbol{\theta}}=\arg\max
_{\boldsymbol{\theta}\in\Theta}f_{\mathbf{t}}(\mathbf{t}(\mathbf{y}%
_{n})|\boldsymbol{\theta})$ using a sufficient summary statistic
$\mathbf{t}(\mathbf{y}_{n})$ of the data. However, this early method is
limited to relatively simple Bayesian modeling scenarios in which
$f_{\mathbf{t}}(\mathbf{t}(\mathbf{y}_{n})|\boldsymbol{\theta})$ can readily
be computed and maximized over $\boldsymbol{\theta}$, whereas this limitation
can be avoided by employing the sampling step
\citep[][pp.2025-6]{BeaumontEtAl02}%
.

R-ABC\ approximates the exact model likelihood $f(y\,|\,\boldsymbol{\theta})$
by the proportion of synthetic data sets that are similar to the observed
data
\citep[][]{DiggleGratton84}%
. For discrete data $\mathbf{y}_{n}$ it may be natural to use no tolerance
($\epsilon=0$) and summary statistics ($\mathbf{t}(\mathbf{y}_{n}%
)\equiv\mathbf{y}_{n}$, $\mathbf{t}(\mathbf{z}_{n})\equiv\mathbf{z}_{n}$)
\citep[][]{TavareEtAl97}%
. For continuous data, some tolerance and summary statistics are more useful
\citep[][]{FuLi97}%
. When $\mathbf{t}$ is sufficient, the R-ABC\ posterior density function
\begin{equation}
\pi_{L}(\boldsymbol{\theta}\,|\,\mathbf{y}_{n})\propto%
{\displaystyle\int}
\mathbf{1}(\rho_{\mathbf{t}(\mathbf{y}_{n}),\mathbf{t}(\mathbf{z}_{n})}%
\leq\epsilon)f(\mathbf{z}_{n}\,|\,\boldsymbol{\theta})\mathrm{d}\mathbf{z}%
_{n}\pi(\boldsymbol{\theta})
\end{equation}
converges to the true posterior $\pi(\boldsymbol{\theta}\,|\,\mathbf{y}%
_{n})\propto f(\mathbf{y}_{n}\,|\,\boldsymbol{\theta})\pi(\boldsymbol{\theta
})$ as $\epsilon\downarrow0$
\citep[][]{BiauCerouGuyader15}%
. Then, $\pi(\boldsymbol{\theta}\,|\,\mathbf{y}_{n})=\pi(\boldsymbol{\theta
}\,|\,\mathbf{t}(\mathbf{y}_{n}))$ for all $\boldsymbol{\theta}$
\citep[][]{Kolmogorov42,LehmannCasella98}%
. Hence the tolerance $\epsilon$ represents model error
\citep[][]{Wilkinson13}%
.

Sampling only one ($N=1$) synthetic data set per IS\ iteration provides a good
trade-off between computational and posterior estimation efficiency. This is
also true for kernel ABC\ (Section 4.2) and R-ABC-MCMC (Section 5)
\citep[][]{BornnEtAl17}%
.

The output of R-ABC\ can greatly depend on the choice of the tuning parameters
$(\rho,\mathbf{t},\epsilon)$. This has motivated developments of other
ABC\ methods that decrease or eliminate tuning parameter dependence, as
described next in Sections 4.2-4.6.

\subsection{Kernel ABC (K-ABC)}

\textit{K-ABC} employs a smooth kernel function $K(\rho)$ without a tolerance
$\epsilon$, resulting in the approximate posterior density:%
\begin{equation}
\pi_{L}(\boldsymbol{\theta}\,|\,\mathbf{y}_{n})\propto%
{\displaystyle\int}
K(\rho_{\mathbf{t}(\mathbf{y}_{n}),\mathbf{t}(\mathbf{z}_{n})})f(\mathbf{z}%
_{n}\,|\,\boldsymbol{\theta})\mathrm{d}\mathbf{z}_{n}\pi(\boldsymbol{\theta}).
\end{equation}
After running the ABC-IS\textit{\ }algorithm, the posterior mean of
$\boldsymbol{\theta}$ is estimated by:
\begin{equation}
\dfrac{%
{\textstyle\sum\nolimits_{s=1}^{S}}
K_{\delta}(||\mathbf{t}(\mathbf{y}_{n})-\mathbf{t}(\mathbf{z}_{n}%
^{(s)})||)\boldsymbol{\theta}_{s}}{%
{\textstyle\sum\nolimits_{s=1}^{S}}
K_{\delta}(||\mathbf{t}(\mathbf{y}_{n})-\mathbf{t}(\mathbf{z}_{n}^{(s)})||)}.
\end{equation}

K-ABC\ may employ the Epanechnikov kernel $K(\rho_{\mathbf{t}(\mathbf{y}%
_{n}),\mathbf{t}(\mathbf{z}_{n})})\equiv K_{\delta}(||\mathbf{t}%
(\mathbf{y}_{n})-\mathbf{t}(\mathbf{z}_{n})||)$ with bandwidth $\delta>0$ and
Euclidean norm $||\mathbf{\cdot}||$
\citep[][]{BeaumontEtAl02}%
, and need not employ summary statistics
\citep[][]{Wilkinson13,TurnerSederberg12,TurnerVanZandt14}%
.

For K-ABC and R-ABC, if the summary statistic(s) obeys a central limit
theorem, then the posterior mean estimator of any parameter function can be
asymptotically normal (as $n\rightarrow\infty$), with mean square error equal
to that of the maximum likelihood estimator based on the summary statistic(s).
Further, for kernel ABC, this estimator is efficient in the sense that the
MC\ error of ABC increases its mean square error only by a factor of
$1+O(1/S)$ for a fixed MC sample size $S$
\citep[][]{LiFearnhead18,DeanSingh11}%
.

\subsection{Coupled ABC (C-ABC)}

C-ABC
\citep[][]{Neal12}
treats ABC as an inverse problem. A c.d.f. $F(y)=\Pr(Y\leq y)$ has a
(generalized) inverse $F^{-1}(u)=\inf_{y}\{y:F(y)\geq u\}$, and $F^{-1}(U)\sim
Y$ if $U\sim U(0,1)$. This means that it is possible to use a deterministic
(coupling) function $Z=f(\boldsymbol{\theta},\boldsymbol{u})$ to map every
realization $\boldsymbol{y}$ of the model likelihood $f(\cdot$\thinspace
$|\,\boldsymbol{\theta})$ to a realization $\boldsymbol{u}\in\lbrack0,1]^{d}$
of the random vector $\boldsymbol{U}$ having p.d.f. $\pi(\mathbf{u})$.
Further, multiple coupled synthetic data sets can be simulated as
$\mathbf{z}_{n}(\boldsymbol{u},\boldsymbol{\theta}_{1}),\mathbf{z}%
_{n}(\boldsymbol{u},\boldsymbol{\theta}_{2}),\ldots,\mathbf{z}_{n}%
(\boldsymbol{u},\boldsymbol{\theta}_{k})$, using the same $\boldsymbol{u}$
over $k$ parameter values, $\boldsymbol{\theta}_{1},\boldsymbol{\theta}%
_{2},\ldots,\boldsymbol{\theta}_{k}$, according to the solution of the inverse
problem based on identifying the set $\Theta_{\mathbf{z}}=\{\boldsymbol{\theta
}:\mathbf{z}=\mathbf{z}_{n}(\boldsymbol{u},\boldsymbol{\theta})\}$, given
$\boldsymbol{u}$. Table 1 shows that C-ABC assumes an approximate likelihood
defined by the kernel $K(\rho_{\mathbf{t}(\mathbf{y}_{n}),\mathbf{t}%
(\mathbf{z}_{n}\{\mathbf{u},\boldsymbol{\theta}\})})$ that is marginalized
over the p.d.f. $\pi(\boldsymbol{u})$.

\subsection{Synthetic Likelihood (SL)\ Methods}

A SL\ method generates multiple ($N\geq1$) samples of synthetic data sets per
sampling iteration. The classical SL\ method assumes the approximate
likelihood to be the multivariate normal p.d.f. $\mathrm{n}(\cdot
\,|\,\boldsymbol{\mu},\mathbf{\Sigma})$, with $\widehat{\boldsymbol{\eta}%
}_{\boldsymbol{\theta},N}^{\{\mathbf{t}(\mathbf{z}_{n(k)})\}}%
=(\widehat{\boldsymbol{\mu}}_{\theta,N},\widehat{\mathbf{\Sigma}%
}_{\boldsymbol{\theta},N})$ the maximum likelihood estimate (MLE) obtained
from $\{\mathbf{t}(\mathbf{z}_{n(k)})\}_{k=1}^{N}$, assuming that the summary
statistics $\mathbf{t}$ are asymptotically normal
\citep[][]{Wood10b,HyrienEtAl05}%
. In terms of the ABC-IS algorithm, iterative step $(a)$ generates a prior
sample $\boldsymbol{\theta}_{s}\sim\pi(\boldsymbol{\theta})$; step $(b)$
estimates $\widehat{\boldsymbol{\eta}}_{\boldsymbol{\theta},N}^{\{\mathbf{t}%
(\mathbf{z}_{n(k)}^{(s)})\}}=(\widehat{\boldsymbol{\mu}}_{\theta,N}%
^{(s)},\widehat{\mathbf{\Sigma}}_{\theta,N}^{(s)})$ from $N$ sampled synthetic
data sets, and step (c)\ weights $\boldsymbol{\theta}_{s}$ by $\omega
_{s}\equiv\mathrm{n}(\mathbf{t}(\mathbf{y}_{n})\,|\,\widehat{\boldsymbol{\mu}%
}_{\theta,N}^{(s)},\widehat{\mathbf{\Sigma}}_{\theta,N}^{(s)})$.

Several other SL\ methods relax the normality assumption, including those that
replace the normal SL by: a GARCH model with time-dependent normal
heteroscedastic variances
\citep[][]{GallantMcCulloch09}%
; a finite mixture of normal experts model with covariate-dependent kernel
densities and mixture weights
\citep[][]{FanEtAl13}%
; a Gaussian process (GP)\ for each individual (marginal) summary statistic,
assuming independent summary statistics
\citep[][]{MeedsWelling14,JabotEtAl14}%
; a GP for the log SL
\citep[][]{Wilkinson14}%
, estimated by sequential history matching
\citep[][]{CraigEtAl97}%
; a GP\ for the discrepancy between observed and simulated synthetic data
\citep[][]{GutmannCorander16}%
, estimated by Bayesian optimization
\citep[][]{Jones01}%
; and other general SLs
\citep[][]{GallantMcCulloch09,DrovandiEtAl11,DrovandiEtAl15,GutmannCorander16}%
.

Other SL\ methods relax normality by instead employing: nonparametric kernel
density estimation
\citep[][]{Blum10,Lee_etal12,TurnerSederberg14,MooresEtAl15,GutmannCorander16}%
; minimum distance estimation
\citep[][]{ForneronNg18}%
; estimation via Hamiltonian dynamics
\citep[][]{MeedsEtAl15}%
; an unbiased estimator of the normal density function
\citep[][]{GhuryeOlkin69}
assuming asymptotically normal summary statistics $\mathbf{t}$; and
MC\ posterior estimation of the normal parameters $(\boldsymbol{\mu}_{\theta
},\mathbf{\Sigma}_{\boldsymbol{\theta}})$
\citep{PriceEtAl18}%
.

\subsection{Empirical Likelihood\ Method (EL-ABC)}

EL-ABC specifies the kernel function by the empirical likelihood
($\mathrm{EL}$),
\begin{equation}
L_{\boldsymbol{\eta}}(\mathbf{y}_{n}\,|\,\boldsymbol{\theta})\equiv
\widehat{L}_{n}^{el}(\mathbf{y}_{n}\,|\,\boldsymbol{\theta})=%
\begin{array}
[c]{c}%
\max\\%
\genfrac{}{}{0pt}{1}{\{\mathbf{p}(\boldsymbol{\theta})\,:\mathbf{\,p}%
\in\lbrack0,1]^{n};}{\mathbb{E}_{F}[h(Y,\boldsymbol{\theta})]=0\}}%
\end{array}
\,%
{\displaystyle\prod\limits_{i=1}^{n}}
p_{i}(\boldsymbol{\theta}), \label{EL}%
\end{equation}
subject to chosen constraints of the form $\mathbb{E}_{F}%
[h(Y,\boldsymbol{\theta})]=0$
\citep[][]{MengersenPudloRobert13}%
. For example, if the model parameter represents the mean, $\boldsymbol{\theta
}=\theta=\mathbb{E}_{f}[Y]$, and $h(y,\theta)=y-\theta$, then the constraint
is $\theta=%
{\textstyle\sum\nolimits_{i=1}^{n}}
p_{i}y_{i}$. The EL\ (\ref{EL}) assumes $n$ i.i.d. observations, but it can
handle dependent observations by reformulating it as a dynamic regression
model that captures underlying iid structure. In terms of the ABC-IS
algorithm, EL-ABC in iterative step $(a)$ draws a prior sample
$\boldsymbol{\theta}_{s}\sim\pi(\boldsymbol{\theta})$; in step (b)\ finds the
EL\ $\widehat{L}_{n}^{el}(\mathbf{y}_{n}\,|\,\boldsymbol{\theta}_{s})$; and in
step (c)\ assigns the IS\ weight $\overline{\omega}_{s}\equiv\widehat{L}%
_{n}^{el}(\mathbf{y}_{n}\,|\,\boldsymbol{\theta}_{s})$.

\subsection{Bootstrap Likelihood Method (BL-ABC)}

BL-ABC
\citep[][]{ZhuMarinLeisen16}
first estimates the exact model likelihood by a kernel density estimate
$L_{\boldsymbol{\eta}}(\mathbf{y}_{n}\,|\,\boldsymbol{\theta})\equiv
\widehat{L}_{n}^{bl}(\mathbf{y}_{n}\,|\,\boldsymbol{\theta})$ of nested
bootstrap samples of a point-estimator $\widehat{\boldsymbol{\theta}}$ of
$\boldsymbol{\theta}$, from the original data $\mathbf{y}_{n}$
\citep[][]{DavisonHinkleyWorton92}%
. The nested bootstrap has two stages. At the first stage, $J$ bootstrap
samples of data sets $\{\mathbf{y}_{n}^{(j)}\}_{j=1}^{J}$ are generated, then
point estimates $\{\widehat{\boldsymbol{\theta}}_{j}^{\ast}%
=\widehat{\boldsymbol{\theta}}(\mathbf{y}_{n}^{(j)})\}_{j=1}^{J}$ are computed
from them (resp.), where each bootstrap sample $\mathbf{y}_{n}^{(j)}$ is
formed by drawing $n$ samples with replacement from the original data
$\mathbf{y}_{n}$, for $j=1,\ldots,J$. At the second stage, for each of
$j=1,\ldots,J$, a total $K$ bootstrap samples of data sets $\{\mathbf{y}%
_{n}^{(j,k)}\}_{k=1}^{K}$ from $\mathbf{y}_{n}^{(j)}$ are generated ($K=1000$
recommended) to yield point estimates $\{\widehat{\boldsymbol{\theta}}%
_{j,k}^{\ast\ast}=\widehat{\boldsymbol{\theta}}(\mathbf{y}_{n}^{(j,k)}%
)\}_{k=1}^{K}$, and the kernel density estimate:%
\begin{equation}
\widehat{L}_{n}^{bl}(\mathbf{y}_{n}\,|\,\boldsymbol{\theta})\equiv
\widehat{f}(\boldsymbol{\theta}\,|\,\widehat{\boldsymbol{\theta}}_{j}^{\ast
})=\dfrac{1}{Ks}%
{\displaystyle\sum\limits_{k=1}^{K}}
\ker%
\genfrac{(}{)}{}{0}{\boldsymbol{\theta}-\widehat{\boldsymbol{\theta}}%
_{j,k}^{\ast\ast}}{h}%
, \label{BL}%
\end{equation}
where $\ker(\cdot)$ is a smooth (e.g., Epanechnikov) kernel with bandwidth
$h>0$. Then the bootstrap likelihood (BL)\ is constructed by fitting a
scatterplot smoother to the $J$\ pairs $\{(\widehat{\boldsymbol{\theta}}%
_{j}^{\ast},\log\widehat{f}(\widehat{\boldsymbol{\theta}}_{n}%
\,|\,\widehat{\boldsymbol{\theta}}_{j}^{\ast}))\}_{j=1}^{J}$. Here,
$\widehat{\boldsymbol{\theta}}_{n}=\widehat{\boldsymbol{\theta}}%
(\mathbf{y}_{n})$, and $\widehat{f}(\widehat{\boldsymbol{\theta}}%
_{n}\,|\,\widehat{\boldsymbol{\theta}}_{j}^{\ast})$ provides an estimate of
the likelihood of $\widehat{\boldsymbol{\theta}}_{n}$ given
$\boldsymbol{\theta}\equiv\widehat{\boldsymbol{\theta}}_{j}^{\ast}$
\citep[][]{DavisonHinkleyWorton92}%
.

For BL-ABC, the ABC-IS algorithm is run with step $(a)$ drawing a prior sample
$\boldsymbol{\theta}_{s}\sim\pi(\boldsymbol{\theta})$; step $(b)$ is skipped;
and step $(c)$\ calculates the importance sampling weight by $\overline
{\omega}_{s}\equiv\widehat{L}_{n}(\mathbf{y}_{n}\,|\,\boldsymbol{\theta}_{s}%
)$. The empirical and bootstrap likelihoods asymptotically agree to order
$n^{-1/2}$, and converge to the true model likelihood $f(\mathbf{y}%
\,|\,\boldsymbol{\theta})$ as $n\rightarrow\infty$
\citep[][]{DavisonHinkleyWorton92}%
.

BL-ABC can be extended in a few ways, owing to the general nature of the
bootstrap method. First, if $\widehat{\boldsymbol{\theta}}-\boldsymbol{\theta
}$ is a pivotal quantity such that $\widehat{\boldsymbol{\theta}%
}-\boldsymbol{\theta}\sim H$, with $H$ not involving $\boldsymbol{\theta}$,
then the bootstrap likelihood estimate $\widehat{L}_{n}^{bl}(\mathbf{y}%
_{n}\,|\,\boldsymbol{\theta})$ can be more simply constructed by the ordinary
single bootstrap
\citep[][]{BoosMonahan86}%
. Second, BL-ABC can be easily extended to handle non-iid dependent data
through the use of the regression residual, parametric, or the pairs
bootstrap
\citep[][]{ZhuMarinLeisen16}%
. Third, BL-ABC can incorporate R-ABC, but at a higher computational cost
\citep[][]{ZhuMarinLeisen16}%
. Fourth, if the model parameter is scalar-valued ($\boldsymbol{\theta}%
\equiv\theta$), BL-ABC can be used to construct an empirical likelihood using
a single bootstrap, without kernel density estimation
\citep[][]{Pawitan00}%
. Finally, BL-ABC can been extended to "$n$ choose $m$" or $m/n$ bootstrap
($m\leq n$) sampling per iteration
\citep[][]{LiangKimSong16}%
.

\section{Combining ABC\ with Other\ Algorithms}

ABC\ methods each have been combined with MCMC (usually MH), SMC, PMC, VI, and
simulated annealing (SA) algorithms, usually in order to speed up posterior computations.

R-ABC-MCMC
\citep
[][]{MarjoramEtAl03,BortotEtAl07,WegmannEtAl09,Lee_etal12,GirolamiEtAl13,LyneEtAl15}%
, R-ABC-SMC
\citep
{SissonEtAl07,RobertEtAl08,ToniEtAl09,PetersFanSisson12,PetersSissonFan12,DelMoralDoucetJasra12,DrovandiPettitt11b,SedkiEtAl12,Filippi13,GolchiCampbell16}%
, R-ABC-PMC
\citep[][]{BeaumontEtAl09,BaragattiEtAl13,Murakami14,IshidaEtAl15}
typically provide automatic tolerance ($\epsilon$) selection. The variance
bounds and geometric ergodicity of R-ABC-MCMC were studied in relation to that
of the MH algorithm for models with intractable likelihood
\citep[][]{LeeLatuszynski14}%
. Also, VI including expected-propogation (EP) methods were proposed to define
K-ABC-EP, R-ABC-VI and K-ABC-VI methods
\citep[][]{BarthelmeChopin11,BarthelmeChopin14,TranEtAl17}%
, and simulated annealing (SA)\ was proposed to define a K-ABC-SA\ method
\citep{AlbertEtAl15}%
.

K-ABC-MCMC
\citep[][]{TurnerVanZandt14}
and C-ABC-MCMC
\citep[][]{NealHuang15,SpenceBlackwell16}
can mix better than R-ABC-MCMC, especially in the tails of the target
posterior distribution
\citep[][]{SissonEtAl07}%
. This may also be true for SL-MCMC
\citep[][]{GallantMcCulloch09,DrovandiEtAl11,PhamEtAl14,Beaumont03}%
, SL-VI\
\citep{OngEtAl18}%
, EL-ABC-MCMC, EL-ABC-PMC, EL-ABC-SMC
\citep[][]{MengersenPudloRobert13}%
, and BL-ABC-MCMC with MH\
\citep[][]{LiangKimSong16}%
.

Random forests
\citep[][]{Breiman01a}
was proposed to estimate the MH\ acceptance ratio as a function of simulated
summary statistics
\citep[][]{PhamEtAl14}%
, based on bootstrap aggregation
\citep[][]{Breiman96}
of the optimal classification predictions of many Classification and
Regression Trees
\citep[CARTs; ][]{BreimanEtAl84}
fitted over resamples of the data $\mathbf{y}_{n}$. Classical SL's\
\citep[][]{Wood10b}
normal synthetic likelihood amounts to assuming the quadratic discriminant
analysis classifier.

\section{ABC\ for Model Choice}

Model choice aims to find the model from a set of $D$ considered models
$\{\mathcal{M}_{d}\}_{d=1}^{D}$ that has the best predictive utility for the
underlying process that generated the given data set, $\mathbf{y}_{n}$. There
are several ABC methods for model choice
\citep[][]{PritchardEtAl99}%
.

One method for R-ABC or K-ABC estimates the posterior model probabilities from
a multinomial logit regression of model indices on summary statistics
$\mathbf{t}(\mathbf{z}_{n})$ sampled from an R-ABC algorithm, locally-weighted
and conditionally on $\mathbf{t}(\mathbf{x}_{n})$
\citep
[][]{FagundesEtAl07,Beaumont08,BlumJakobsson10,EstoupEtAl12,PrangleEtAl14}%
. A method for R-ABC
\citep[][]{FrancoisLaval11}
approximates each model's deviance information criterion
\citep[][]{SpiegelhalterBestCarlinvanderLinde02}
by using kernel density estimation (tolerance $\epsilon$ bandwidth) of the
deviance statistic, obtained from summary statistics of posterior predictive
samples generated per sampling iteration. Another R-ABC method
\citep[][]{Kobayashi14}
employs an MH\ algorithm that proposes jumps between models, based on the
pseudo-marginal approach that handles intractable likelihoods
\citep[][]{AndrieuRoberts09}%
.

A general ABC\ model choice method employs random forests
\citep[][]{PudloEtAl16}%
, as follows. First, a prior probability $\pi(\mathcal{M}_{d})$ is assigned to
each model $\mathcal{M}_{d}$ with intractable likelihood $f_{d}(\mathbf{z}%
_{n}\,|\,\boldsymbol{\theta}_{d})$ and prior $\pi_{d}(\boldsymbol{\theta}%
_{d})$, for a given set $\{\mathcal{M}_{d}\}_{d=1}^{D}$ of compared models.
Then a table is constructed by taking $N_{ref}$ samples of $(d,\mathbf{t}%
(\mathbf{z}_{n}))$ from $\pi(\mathcal{M}_{d})\pi_{d}(\boldsymbol{\theta}%
_{d})f_{d}(\mathbf{z}_{n}\,|\,\boldsymbol{\theta}_{d})$. Next, the method fits
a CART $T_{b}$ that classifies model indices with $\mathbf{t}(\mathbf{z}_{n}%
)$, to each of $B$ bootstrap samples (with replacement)\ of size
$N_{boot}<N_{ref}$ from the table. Finally, the single best predictive model
from $\{\mathcal{M}_{d}\}_{d=1}^{D}$ is identified as the one with the model
index that receives the most votes (optimal classifications) from the $B$
fitted CARTs.

Finally, R-ABC, R-ABC-MCMC (MH), and R-ABC-SMC algorithms can be extended for
assessing the fit of a single model to data
\citep[][]{RatmannEtAl11,RatmannEtAl09,RobertEtAl10,RatmannEtAl10}%
.

\section{Software Packages for ABC}

There are at least 28 statistical software packages that support many of the
ABC methods that were mentioned earlier
\citep[in part, from][]{NunesPrangle15,KousathanasEtAl17}%
. Table 2 provides a summary of these packages. Nearly all packages are based
on R-ABC, while a few recent packages handle K-ABC or SL-ABC. Also, 9 of these
packages can be applied to general models, while all the other packages
address specific models, scientific fields, or focuses on model selection tasks.

According to Table 2 and the review of applications given in Section 1, R-ABC
seems to be most widely applied in the population genetics, phylogeography,
systems biology, archaeology, and cosmology fields. However, in the future it
is expected that fields will more frequently apply the more recent ABC methods
since they have some advantages over R-ABC.%

\begin{table}[tbp] \centering
\begin{tabular}
[c]{lcll}%
\textbf{ABC\ Package} & \textbf{Author(s)} & \textbf{ABC\ method(s)} &
\textbf{Models/Field}\\\hline
\texttt{abc*} &
\citep[][]{CsilleryEtAl12}%
& R-ABC & General\\
\texttt{abc\_distrib} &
\citep[][]{BeaumontEtAl02}%
& R-ABC & General\\
\texttt{abc\_nnet} &
\citep[][]{BlumFrancois10}%
& R-ABC & General\\
\texttt{ABCreg} &
\citep[][]{Thornton09}%
& R-ABC & General\\
\texttt{ABCtoolbox*} &
\citep[][]{WegmannEtAl10}%
& R-ABC, R-ABC-MCMC, & General\\
&  & R-ABC-PMC & General\\
\texttt{abctools} &
\citep[][]{NunesPrangle15}%
& R-ABC & \\
\texttt{ABrox*} &
\citep[][]{MertensEtAl18}%
& R-ABC, R-ABC-MCMC, & Model selection focus\\
\texttt{DREAM*} &
\citep[][]{Vrugt16}%
& K-ABC-MCMC & General\\
\texttt{EasyABC} &
\citep[][]{JabotEtAl13}%
& R-ABC, R-ABC-SMC, & General\\
&  & SL-ABC, K-ABC-SA & \\
\texttt{synlik} &
\citep[][]{FasioloWood14}%
& SL-ABC & General\\\hline
\texttt{ABC-EP} &
\citep[][]{BarthelmeChopin14}%
& R-ABC-EP & State space (\& related)\\
\texttt{abc-sde} &
\citep[][]{Picchini13}%
& R-ABC-MCMC & SDE\\
\texttt{gk} &
\citep{Prangle17}%
& R-ABC & $g$-and-$k$ (or $h$) models\\\hline
\texttt{2BAD*} &
\citep[][]{BrayEtAl10}%
& R-ABC & Population genetics\\
\texttt{ABC4F} &
\citep[][]{FollEtAl08}%
& R-ABC & Population genetics\\
\texttt{Bayes-SSC} &
\citep[][]{AndersonEtAl05}%
& R-ABC & Population genetics\\
\texttt{DIY-ABC*} &
\citep{CornuetEtAl08}%
& R-ABC & Population genetics\\
\texttt{msABC} &
\citep[][]{PavlidisEtAl10}%
& R-ABC & Population genetics\\
\texttt{onesamp} &
\citep[][]{TallmonEtAl08}%
& R-ABC & Population genetics\\
\texttt{PopABC} &
\citep[][]{LopesEtAl09}%
& R-ABC & Population genetics\\
\texttt{REJECTOR*} &
\citep[][]{JobinMountain08}%
& R-ABC & Population genetics\\
\texttt{msBayes} &
\citep[][]{HickersonEtAl07}%
& R-ABC & Phylogeography\\
\texttt{MTML-msBayes*} &
\citep[][]{HuangEtAl11}%
& R-ABC & Phylogeography\\\hline
\texttt{ABC-SysBio*} &
\citep[][]{LiepeEtAl10}%
& R-ABC, R-ABC-SMC & Systems biology\\\hline
\texttt{WARN} &
\citep[][]{TsutayaYoneda13}%
& R-ABC & Archaeology\\\hline
\texttt{abcpmc} &
\citep[][]{AkeretEtAl15}%
& R-ABC-PMC & Cosmology\\
\texttt{astroABC} &
\citep[][]{JenningsMadigan17}%
& R-ABC-SMC & Cosmology\\
\texttt{CosmoPMC*} &
\citep[][]{KilbingerEtAl11}%
& R-ABC-PMC & Cosmology\\\hline
\end{tabular}
\caption{General and specific-purpose ABC software packages, including their methodological capabilities. An asterisk (*) indicates an ABC package that provides model selection. (SDE = Stochastic Differential Equation.}\label{Table2}%
\end{table}%

\section{Open Problems in\ ABC}

The six types of ABC methods have proven useful for many applications of data
analysis. However, they are not fully satisfactory for reasons that are
summarized in Table 3.\ Table 3 defines the current open problems with ABC,
described in the following subsections.

\noindent%
\begin{table}[tbp] \centering
\begin{tabular}
[c]{|ccccc|}\hline\hline
\multicolumn{1}{||c}{} & $\text{Inferences depend}$ &
\multicolumn{2}{c}{Computational issues?} & \multicolumn{1}{c||}{Provides}\\
\multicolumn{1}{||c}{$%
\begin{array}
[c]{c}%
\text{ ABC}\\
\text{ Method}%
\end{array}
$} & $%
\begin{array}
[c]{c}%
\text{ on tuning}\\
\text{or estimator?}%
\end{array}
$ & $%
\begin{array}
[c]{c}%
\text{Curse of}\\
\text{dimension?}%
\end{array}
$ & $%
\begin{array}
[c]{c}%
\text{due to }\\
\text{big }n\text{?}%
\end{array}
$ & \multicolumn{1}{c||}{$%
\begin{array}
[c]{c}%
\text{model}\\
\text{choice?}%
\end{array}
$}\\\hline\hline
$\text{Rejection}$ & \multicolumn{1}{|c}{$\rho,\mathbf{t},\epsilon,N$} &
\multicolumn{1}{|c}{{\small In tuning}} & \multicolumn{1}{|c}{{\small Slow}} &
\multicolumn{1}{|c|}{{\small Posterior model}}\\\cline{1-1}\cline{1-2}%
\cline{2-2}%
$\text{Kernel}$ & \multicolumn{1}{|c}{$\rho,N,$ {\small maybe} $\mathbf{t}$} &
\multicolumn{1}{|c}{{\small parameter}} &
\multicolumn{1}{|c}{{\small sampling}} &
\multicolumn{1}{|c|}{{\small probabilities from}}\\\cline{1-1}\cline{1-2}%
\cline{2-2}%
$\text{Coupled}$ & \multicolumn{1}{|c}{$\rho,\mathbf{t},\epsilon,N$} &
\multicolumn{1}{|c}{$(\rho,\mathbf{t},\epsilon,N)$} &
\multicolumn{1}{|c}{${\small N}$ {\small synthetic}} &
\multicolumn{1}{|c|}{{\small regression of}}\\\cline{1-2}%
$\text{SL}$ & \multicolumn{1}{|c}{$\mathbf{t},N$} &
\multicolumn{1}{|c}{{\small selection}} & \multicolumn{1}{|c}{{\small data
sets}} & \multicolumn{1}{|c|}{{\small model indices on }$\mathbf{t}$}\\\hline
$\text{EL}$ & \multicolumn{1}{|c}{{\small Needs estimator}} &
\multicolumn{1}{|c}{{\small Parameter}} & \multicolumn{1}{|c}{{\small Slow
estim.,}} & \multicolumn{1}{|c|}{No}\\\cline{1-1}\cline{1-1}\cline{5-5}%
$\text{BL}$ & \multicolumn{1}{|c}{{\small of function of }$\boldsymbol{\theta
}$} & \multicolumn{1}{|c}{{\small estimation}} &
\multicolumn{1}{|c}{{\small resampling.}} & \multicolumn{1}{|c|}{No}\\\hline
\end{tabular}
\caption{The open problems in ABC, including the computational complexity of each ABC method.}%
\end{table}%

The table conveys that several aspects can together contribute to the
computational complexity of each ABC method, such as the sample size or the
number of model parameters or variables. However, these issues may potentially
be less important as increasing computing power continues to become more
available
\citep{SunnaakerEtAl13}%
.

\subsection{Tuning Parameter $(\rho,\mathbf{t},\epsilon,N)$ and
Point-Estimator Dependence}

Posterior inferences from R-ABC, K-ABC, C-ABC, and SL-ABC can be very
sensitive to the choice of tuning parameters $(\rho,\mathbf{t},\epsilon,N)$
\citep[e.g.,][]{MarinEtAl12}%
.

A typical choice of $\rho$ is the squared distance. But posterior inferences
differ across other reasonable choices of distance measures, and there does
not seem to be one "best" distance measure for general ABC practice. K-ABC
employs a smooth kernel function $K_{\delta}(\rho)$ of $\rho$. The kernel
bandwidth needs to be carefully chosen, perhaps by matching the resulting
posterior credible intervals with the correct coverage levels
\citep[][]{PrangleEtAl14b}%
.

When employed, summary statistics $\mathbf{t}$ should be sufficient. But
sufficient statistics are not available from many Bayesian models and data
sets. Insufficient summary statistics when used should be carefully selected
to ensure accurate posterior inferences
\citep[e.g.,][]{DidelotEtAl11,MarinEtAl14,RobertEtAl11}%
. Many statistics selection methods have been developed
\citep
[][]{JoyceMarjoram08,WegmannEtAl09,NunesBalding10,FearnheadPrangle12,AeschbacherEtAl,BlumEtAl13,GleimPigorsch13,JiangEtAl15,RuliEtAl16,CreelKristensen16,MitrovicEtAl16}%
.

In R-ABC,\ the tolerance $\epsilon$ represents model approximation error
\citep[][]{Wilkinson13}
and controls the trade-off between computational speed and estimation
accuracy. Running a R-ABC\ algorithm with overly-small $\epsilon$ requires
many model evaluations due to the large rejection probability, making it
time-consuming to get a large number of acceptances. When $\epsilon$ is too
large, the algorithm poorly approximates and overstates uncertainty in the
posterior distribution. Several proposed solutions include selecting the
tolerance from the MC\ sampling run
\citep
[][]{SissonEtAl07,BeaumontEtAl09,ToniEtAl09,DelMoralDoucetJasra12,SilkEtAl12,LenormandEtAl13,ChiachioEtAl14}
or from posterior asymptotics
\citep[][]{BarberEtAl15}%
; and running a R-ABC algorithm for each of the $n$ components of a factorized
exact model likelihood via the Markov property (if factorizable), possibly
with lower tolerance and computational cost and without $(\rho,\mathbf{t})$
\citep[][]{WhiteKypraiosPreston15,BarthelmeChopin11,BazinEtAl10}%
.

In SL-ABC, $N$ controls the trade-off between computational speed and
estimation accuracy
\citep[e.g.,][]{DrovandiEtAl15}%
, and inferences may be highly-sensitive to $N$ when the SL\ is not smooth
\citep[][Section 4]{GutmannCorander16}%
. Recent extensions of SL-ABC can decrease this sensitivity
\citep{PriceEtAl18}%
.

Some issues are avoided by employing ABC methods that do not rely on all
tuning parameters $(\rho,\mathbf{t},\epsilon)$. K-ABC and C-ABC\ do not use
$\epsilon$ and may not require $\mathbf{t}$, and SL-ABC does not use
$(\rho,\epsilon)$, but they depend on other tuning parameters. C-ABC\ requires
solving the inverse problem, which is not possible for all model likelihoods.
For R-ABC, a method can select $(\rho,\mathbf{t},\epsilon)$
\citep[][]{RatmannEtAl13}%
. The factorization method does not require $(\rho,\mathbf{t})$.

In summary, there are many proposed solutions to address tuning parameter
dependence in ABC, but none appear to be neat and fully satisfactory.

EL-ABC\ and BL-ABC do not employ any tuning parameters. But they still require
a point estimator for $\mathbf{p}(\boldsymbol{\theta})$ and
$\boldsymbol{\theta}$ (resp.), which may not be available for the given
Bayesian model at hand. In EL-ABC it can be difficult to specify the moment
constraints needed to estimate $\mathbf{p}(\boldsymbol{\theta})$. See Sections
8.2-8.3 for related issues.

\subsection{The Curse of Dimensionality}

The curse of dimensionality
\citep[][]{Bellman61}
can pose challenges to the tuning parameters. In R-ABC, K-ABC, and C-ABC,
squared distance is a typical choice for $\rho$, but it suffers from the curse
beyond three dimensions
\citep[e.g.,][]{HolmesMallick03}%
, perhaps also true for general distance measures of summary statistics
\citep[][]{BlumEtAl13}%
. Even when sufficient statistics are available, they often have the same
dimensionality as the sample size. For instance, while $\dim(\mathbf{t}%
)\geq\dim(\boldsymbol{\theta})$ is required for parameter identifiability,
this implies that high-dimensional summary statistics $\mathbf{t}$ are
required to adequately describe the information of a high-dimensional
$\boldsymbol{\theta}$. It is not uncommon in ABC\ practice that $\dim
(\mathbf{t})\gg\dim(\boldsymbol{\theta})$
\citep[][]{AllinghamEtAl09,BortotEtAl07}%
. Otherwise, low-dimensional sufficient statistics may be computationally intractable.

Proposed solutions include: performing a (linear, local, or nonlinear)
regression of accepted proposed parameter values onto the distance between the
simulated and observed summary statistics, and then estimating the posterior
distribution by sampling the regression error distribution conditionally on
zero distance
\citep
[][]{BeaumontEtAl02,Thornton09,Blum10,BlumFrancois10,NottEtAl14,GutmannCorander16}%
; performing regression adjustments on low-dimensional marginal posterior
distributions and summary statistics
\citep[][]{NottEtAl14,LiEtAl17}%
, since low-dimensional posteriors can often be estimated well; methods to
reduce the dimensionality of the summary statistics
\citep[][]{BlumEtAl13}%
; and the factorized likelihood method using no tuning parameters
$(\rho,\mathbf{t})$
\citep[][]{WhiteKypraiosPreston15}%
.

A high-dimensional model parameter $\boldsymbol{\theta}$ also affects EL-ABC
and BL-ABC by making point estimation of $\mathbf{p}(\boldsymbol{\theta})$ and
$\boldsymbol{\theta}$ (resp.) more challenging and time consuming, especially
when the data sample size ($n$) is large. Then, in EL-ABC the constraints for
$\mathbf{p}(\boldsymbol{\theta})$ are difficult to specify
\citep[][]{ZhuMarinLeisen16}%
, and in BL-ABC a large number of bootstrap samples are needed for reliable
kernel estimation of the approximate likelihood. Low-dimensional summary
statistics may help.

The ABC-PaSS algorithm
\citep[][]{KousathanasEtAl16}
(i.e., ABC for Parameter Specific Statistics ABC) was designed to tackle the
problem of high-dimensionality that arises frequently in inference for complex
models, which are typical in biology and population genetics. This approach is
based on Gibbs sampling, where each parameter in the model is updated
independently, and this update is accepted or rejected based on the Euclidean
distance of parameter-specific statistics to the observed data.

\subsection{Large Sample Size ($n$) Issues}

For R-ABC, K-ABC, and SL-ABC, simulating $N\geq1$\ synthetic data sets of size
$n$ per sampling iteration can be computationally costly, especially when the
data sample size $n$\ is large. For certain parameter space regions, a small
number of simulations $N$ may suffice to conclude that the approximate
likelihood cannot take on a significant value. Proposed solutions to such
issues include an extension of SL-ABC that uses Bayesian optimization for
estimating SL\ parameters instead of simulating synthetic data
\citep[][]{GutmannCorander16}%
; and an R-ABC IS\ algorithm that saves time by sometimes abandoning the
simulation of synthetic data sets that likely poorly match the observed data
\citep[][]{Prangle16}%
.

\textit{Approximate ABC}\ (AABC)\ can provide a faster R-ABC method to
simulate synthetic data sets in settings where it is too computationally
costly to directly simulate from the exact model likelihood
\citep[][]{BuzbasRosenberg15}%
. AABC\ initializes the ABC-IS algorithm by drawing a small number $N^{\ast}$
of prior samples $\{\boldsymbol{\theta}^{\ast}\}_{k=1}^{N^{\ast}%
}\overset{\text{iid}}{\sim}\pi(\boldsymbol{\theta})$ and synthetic data sets
$\mathbf{Z}^{\ast}=(\mathbf{z}_{n(k)}^{\ast})_{k=1}^{N^{\ast}}%
\overset{\text{iid}}{\sim}f(\mathbf{\cdot}\,|\,\boldsymbol{\theta})$. Then
after drawing $\boldsymbol{\theta}_{s}$ in iterative step $(a)$, AABC in
iterative step $(b)$\ draws $\boldsymbol{\phi}$ from a Dirichlet distribution
with precision parameters given by the Epanechnikov kernel weights of
$\{\boldsymbol{\theta}^{\ast}\}_{k=1}^{N^{\ast}}$ centered at
$\boldsymbol{\theta}_{s}$, and samples a synthetic data set $\mathbf{z}%
_{n}=(z_{i})_{i=1}^{n}$ by drawing $z_{i}=z_{j}^{\ast}$ with probability
$\phi_{j}$ for $i=1,\ldots,n$.

Large sample size ($n$)\ also slows point-estimation. This seems to be true
for SL-ABC extensions that estimate a GP\ per sampling iteration because this
can involve multiple inversions of $n\times n$ matrices, and true for EL-ABC
and BL-ABC which estimate $\mathbf{p}(\boldsymbol{\theta})$ and
$\boldsymbol{\theta}$ (resp.) in each iteration.

\subsection{ABC\ Model Choice Issues}

All current ABC\ model choice methods employ summary statistics $\mathbf{t}$,
and also a tolerance $\epsilon>0$ in some cases. Coherent R-ABC model choice
requires sufficient summary statistics
\citep
[][]{GrelaudEtAl09,DidelotEtAl11,RobertEtAl11,BurrSkurikhin13,MarinEtAl14,CabrasEtAl15}%
. The results of ABC\ model choice can be inconsistent when insufficient
summary statistics are employed
\citep[][]{RobertEtAl11}%
, and can depend heavily on the information amount in the given data set
\citep[][]{StocksEtAl14}%
. But as mentioned, sufficiency is not provided by many models and data sets,
and tolerance selection is not trivial.

Nearly all of the ABC\ model choice methods assign a prior distribution
$\pi(\mathcal{M}_{d})$ on the model space. This implicitly assumes the
$\mathcal{M}$-closed view of model choice, namely that the true model that
generated the data $\mathbf{y}_{n}$ is in the set of models $\{\mathcal{M}%
_{d}\}_{d=1}^{D}$ under consideration
\citep[][]{BernardoSmith94}%
. This view is not uncontroversial because the true model is often unavailable
in practice, which anyway may not accurately predict future data given the
available evidence $\mathbf{y}_{n}$
\citep[][]{VehtariOjanen12}%
. Alternatively, an $\mathcal{M}$-open view may be adopted, avoiding the
explicit construction of an actual true belief model. In this case, model
choice can proceed by comparing models by posterior predictive performance of
future data. This can be done under minimal modeling assumptions by employing
sample re-use cross-validation methods to approximate the future data distribution.

\section{Conclusions}

ABC provides useful inferential methods for Bayesian models with intractable
likelihoods. On the basis of approximate likelihood theory, this article
provided a unifying review of ABC methods in the huge related literature. This
allowed for the methods to be concisely described, classified, and directly
compared, with the aim of promoting future use of ABC methods among
scientists. The review also informed a summary of some of the current open
problems with ABC methods. The review as a whole suggests some future research
directions in ABC, possibly by combining some of the virtues of K-ABC, SL-ABC,
and BL-ABC, while avoiding some of their issues.

\section{Acknowledgements}

This jointly-authored article is supported by NSF grant SES-1156372 to
Karabatsos, and by the European Community's Seventh Framework Programme
[FP7/2007-2013] grant 630677 to Leisen. The authors thank the Editor and two
reviewers for helpful comments on the previous version of this manuscript
(dated July 27, 2017).

\bibliographystyle{imsart-nameyear}
\bibliography{Karabatsos}

\begin{center}

\end{center}

\end{document}